\documentclass[useAMS,usenatbib,referee]{mn2e}
\usepackage{graphicx}

\newcommand{\bea}{\begin{eqnarray}}
\newcommand{\eea}{\end{eqnarray}}
\newcommand{\be}{\begin{equation}}
\newcommand{\ee}{\end{equation}}

\title[The magnetisation of protoplanetary disks]{The 
magnetisation of protoplanetary disks}
\author[G.T. Birk, H. Wiechen, A. Kopp and H. Lesch]{G.T. 
Birk$^{1, 2}$\thanks{E-mail:
birk@usm.uni-muenchen.de}, H. Wiechen$^{2}$, A. 
Kopp$^{3}$ and H. Lesch$^{1, 2}$\\
$^{1}$ Institut f\"ur Astronomie und Astrophysik,
Ludwig-Maximilians-Universit\"at M\"unchen,
Germany\\ 
$^{2}$Centre of Interdisciplinary Plasma Science, 
Garching, Germany\\
$^{3}$Theoretische Physik IV,
Ruhr-Universit\"at Bochum, Bochum, Germany}
\begin{document}
\date{}
\pagerange{\pageref{firstpage}--\pageref{lastpage}} 
\pubyear{2002}

\maketitle

\label{firstpage}

\begin{abstract}
The remanent magnetisation of meteorite material in the 
solar system
indicates that magnetic fields of several Gauss are 
present
in the protoplanetary disk. It is shown that such 
relatively
strong magnetic fields can be generated in dusty 
protoplanetary disks 
by relative shear motions of the charged dust and the 
neutral 
gas components. Self-consistent multi-fluid simulations 
show
that for typical plasma parameters shear flows with
collisional momentum transfer between the different 
components 
of the dusty plasma result in a very efficient generation 
of 
magnetic fields with strengths of about 0.1-1 Gauss on 
spatial scales of 
two astronomical units in about one year. Based on the 
simulations
one may predict that future measurements will reveal the 
strong magnetisation
of circumstellar disks around young stellar objects. 
\end{abstract}
\section{Introduction}

Meteors are commonly considered to be relics of the 
protoplanetary nebula
in our early solar system.
Of particular importance in the context of the early 
evolution of the solar 
system is the remanent magnetisation of meteorite 
material that indicates the
presence of relatively strong magnetic fields ($B=1-10$ 
G) 
in protoplanetary nebula \citep{lev78, sug79}.
The formation of meteorite chondrules may also demand for 
magnetic fields
as the relevant energy source 
(Sonett 1979; Levy \& Araki 1989).
Thus, one faces the question of the origin of such strong 
magnetic fields in protoplanetary disks.
Recent numerical calculations of collapsing magnetised molecular cloud cores
have proven that ambipolar diffusion can solve the magnetic flux loss
problem of the formation of protostellar objects \citep{cio98, des01}.
The magnetic flux can diffuse to the region around the protostellar object
and thus result in a magnetic field of the protoplanetary disk
of some 0.1 G \citep{des01}. The fate of this field
after the stellar object has fully evolved and accretion flow in the disk
sets in is, however, not clear.
In this contribution we consider disk magnetisation in a later stage of 
the stellar evolution, namely after a stellar object with a dipole 
magnetic field has formed.  

A characteristic feature of partially ionised plasmas is 
the possibility that magnetic fields can be generated without any seed 
fields. 
As shown by
(Huba \& Fedder 1993)
a relative plasma-neutral gas flow with non-vanishing 
vorticity results
inevitably in the generation of magnetic fields. This 
effect is described
by the generalised Ohm's law on the grounds of 
a multi-fluid description. The magnetisation
mechanism works significantly more efficiently in a dusty 
plasma
(Birk et al. 1996; Birk et al. 2001)
where the inertia effects and collisional 
momentum transfer are dominated by the heavy dust grains. 

In this contribution we show by means of multi-fluid 
simulations that relative
shear flows lead to a very fast three-dimensional 
magnetic field self-generation in dusty protoplanetary 
disks. 
Magnetic field strengths up to one Gauss are generated on 
the time scale 
of 1yr on spatial scales of some astronomical units (AU).

\section{Self-magnetisation of protoplanetary accretion 
disks}

We consider the following balance equations that govern 
the 
low-frequency macroscopic dynamics of a partially ionised
dusty plasma (see Birk et al. 1996; Birk et al. 2001)
\be
{\partial \rho_d \over \partial t} = -\nabla\cdot (\rho_d {\bf
v_d})\label{1}
\ee
\be
{\partial \rho_i \over \partial t}=
   -\nabla\cdot\Bigl(\rho_i {\bf v_i}\Bigr) \label{2}
\ee
\be
{\partial \rho_n \over \partial t} = -\nabla\cdot (\rho_n {\bf v_n}) \label{3}
\ee
\bea
{\partial (\rho_d{\bf v_d}) \over \partial t}
  =  &-&\nabla\cdot(\rho_d {\bf v_d v_d})
      -\nabla(p_e+p_i+p_d) + (\rho_i + \rho_d) {\bf g}
        \nonumber\\
    &+&(\nabla\times{\bf B})\times{\bf B}
      -\nu_{dn}\rho_d({\bf v_d}-{\bf v_n})\nonumber\\
    &-& \nu_{in}\rho_i \Bigl({\bf v_i}-{\bf v_n}\Bigr)
    - \nu_{en}\rho_e \Bigl({\bf v_e}-{\bf v_n}\Bigr)
\label{4}
\eea
\bea
{\partial (\rho_n {\bf v_n}) \over \partial t}
  =  &-& \nabla\cdot(\rho_n {\bf v_n v_n})
      -\nabla p_n + \rho_n {\bf g} \nonumber\\
     &+&\nu_{dn}\rho_d({\bf v_d}-{\bf v_n})
      +\nu_{in}\rho_i\Bigl({\bf v_i}-{\bf v_n}\Bigr)
     \nonumber\\
     &+&\nu_{en}\rho_e\Bigl({\bf v_e}-{\bf v_n}\Bigr)
\label{5}
\eea
\bea
{1\over \gamma_e-1}{\partial p_e \over \partial t}= 
         &-&{1\over \gamma_e-1} \nabla\cdot (
            p_e{\bf v_e})
       - p_e\nabla\cdot{\bf v_e} + 
    \rho_n\nu_{ne} ({\bf v_e}-{\bf v_n})^2
       \nonumber\\
            &-&2{\rho_d \nu_{de} \over m_d}\biggl({k_B T_e\over \gamma_e-1}
               -{k_B T_d \over \gamma_d-1} \biggr) 
             - 2{\rho_i \nu_{ie} \over m_i}\biggl({k_B T_e\over \gamma_e-1}
               -{k_B T_i \over \gamma_i-1} \biggr) \nonumber\\
            &-&2{\rho_n \nu_{ne} \over m_n}\biggl({k_B T_e\over \gamma_e-1}
               -{k_B T_n \over \gamma_n-1} \biggr) 
      + \rho_d \nu_{de} \Bigl({\bf v_e}-{\bf v_d}\Bigr)^2
      \nonumber\\
     &+&\rho_i \nu_{ie} ({\bf v_e}-
                             {\bf v_i}
        )^2\\
{1\over \gamma_i-1}{\partial p_i \over \partial t} =
         &-&{1\over \gamma_i-1} \nabla\cdot (p_i{\bf v_i}
        )
       - p_i\nabla\cdot{\bf v_i} 
    + {m_n\over m_i+m_n}\rho_i\nu_{in} ({\bf v_i}-{\bf v_n})^2
    \nonumber\\
    &+&{m_d\over m_d+m_i}\rho_i\nu_{id} ({\bf v_i} 
     -{\bf v_d})^2 - 2{\rho_i\nu_{in}\over m_i +m_n}
       \biggl({k_B T_i\over \gamma_i-1}-{k_B T_n\over \gamma_n-1}\biggr)
    \nonumber\\
       &-&2{\rho_i\nu_{id}\over m_i +m_d}
       \biggl({k_B T_i\over \gamma_i-1}-{k_B T_d\over \gamma_d-1}\biggr)
             - 2{\rho_i \nu_{ie} \over m_i}\biggl({k_B T_i\over \gamma_i-1}
               -{k_B T_e \over \gamma_e-1} \biggr) \nonumber\\
     &+&\frac{m_e}{m_i}\rho_i \nu_{ie} ({\bf v_i}-
                             {\bf v_e}
        )^2\\
{1\over \gamma_d-1}{\partial p_d \over \partial t}=
          &-&{1\over \gamma_d-1}\nabla\cdot(p_d{\bf v_d})-p_d\nabla
             \cdot{\bf v_d}
           + {m_e\over m_d}\rho_d\nu_{de}
             ({\bf v_d}-{\bf v_e})^2\nonumber\\
          &+&{m_i\over m_i+m_d}\rho_d\nu_{di}
             ({\bf v_d}-{\bf v_i})^2
          +{m_n\over m_n+m_d}\rho_d\nu_{dn}({\bf v_d}-{\bf v_n})^2\nonumber\\
          &-&2{\rho_d\nu_{di}\over m_i +m_d}
             \biggl({k_B T_d\over \gamma_d-1}-{k_B T_i\over \gamma_i-1}\biggr)
          -2{\rho_d\nu_{dn}\over m_n +m_d}
             \biggl({k_B T_d\over \gamma_d-1}-{k_B T_n\over
               \gamma_n-1}\biggr)\nonumber\\
              &-&2{\rho_d\nu_{de}\over m_d}
             \biggl({k_B T_d\over \gamma_d-1}-{k_B T_e\over
               \gamma_e-1}\biggr) \label{6}
\eea
\bea
{1\over \gamma_n-1}{\partial p_n \over \partial t}=
         &-&{1\over \gamma_n-1}\nabla\cdot(p_n{\bf v_n})-p_n\nabla\cdot
         {\bf v_n}
         \nonumber\\
        &+&{m_i\over m_i+m_n} \rho_n\nu_{ni} ({\bf v_i}\mathord
                                          -{\bf v_n})^2
        +{m_e\over m_n} \rho_n\nu_{ne} \Bigl({\bf v_e}\mathord
                                          -{\bf v_n}\Bigr)^2
        \nonumber\\
             &+&{m_d\over m_n+m_d}\rho_n\nu_{nd}({\bf v_n}-{\bf v_d})^2
                -2{\rho_n\nu_{ni}\over m_i +m_n}
       \biggl({k_B T_n\over \gamma_n-1}-{k_B T_i\over \gamma_i-1}\biggr)
       \nonumber\\
       &-&2{\rho_n\nu_{nd}\over m_n +m_d}
             \biggl({k_B T_n\over \gamma_n-1}-{k_B T_d\over
          \gamma_d-1}\biggr) 
       \nonumber\\
       &-&2{\rho_n\nu_{ne}\over m_n}
             \biggl({k_B T_n\over \gamma_n-1}-{k_B T_e\over
          \gamma_e-1}\biggr) 
       \label{7}
\eea
\bea
{\partial {\bf B} \over \partial t}=
   &-&{m_i c\over e}\nabla\times\biggl({\nabla 
p_i\over\rho_i}\biggr)
    +{{m_i} \over {m_d}} z_d 
\nabla\times\biggl({{\rho_d}\over{\rho_i}}
{\bf v_d}
    \times{\bf B}\biggr) \nonumber \\
   &+&{m_i c\over {4 \pi 
e}}\nabla\times\biggl(\biggl({{\nabla 
\times {\bf B}} \over {\rho_i}}\biggr) \times {\bf B} 
\biggr) - \eta \Delta 
{\bf B} \nonumber\\
   &-&{m_i c\over e}\nabla\times 
     \biggl[{{n_d} \over {n_i}} \biggl(\biggl(z_d - 
{{n_i} \over {n_d}} 
\biggr) {\nu_{id}} + z_d \nu_{in} \biggr){\bf v}_d - 
\nu_{in} {\bf v}_n 
\biggr]
\label{8}     
\eea
where $\rho_{\alpha}$, ${\bf v}_{\alpha}$, $m_\alpha$, $p_\alpha$ and
${\bf B}$ are the 
respective ($\alpha = d, i, e, n$) mass density, bulk velocity, 
particle mass (the ions are assumed to be protons), thermal pressure
($p_\alpha=\rho_\alpha k_B T_\alpha/m_\alpha$, 
where $k_B$ is the Boltzmann constant and $T_\alpha$ is the temperature)
and magnetic field strength. The adiabatic ratios of specific heats
are chosen as $\gamma_\alpha=5/3$.
The collision frequencies for collisions 
(elastic collisions as well as
charge exchange) between species $\alpha$ and $\beta$ are denoted by
$\nu_{\alpha \beta}$ and satisfy $\rho_\alpha\nu_{\alpha\beta}=
\rho_\beta\nu_{\beta\alpha}$.
By $\eta$ the
collisional diffusivity ($\eta= m_i c^2(\nu_{id} + 
\nu_{in})/4\pi n_ie^2$) is denoted.
Ionisation and recombination are neglected in the present context. 
The electron density is calculated from the quasi-neutrality condition
\hbox{($\rho_i/m_i-\rho_e/m_e-z_d\rho_d/m_d=0$)} where $z_d$ is the
negative dust charge number.  
The momentum transfer equations that include external gravitational forces 
with 
the acceleration ${\bf g}$ are formulated as the sum of the balance 
equations of all 
charged components under the assumption of inertialess
ions and electrons (as compared to the massive dust grains).
Most important for the self-generation of magnetic fields 
is the induction 
equation (see also Ciolek \& Mouschovias, 1993) which may 
be derived from the inertialess ion momentum equation. 
The final term in Eq.~(10) is the relevant magnetic field 
self-generation term.
The mass, momentum and pressure balance equations as well 
as the induction equation are integrated numerically by 
the DENISIS code (Schr\"oer et al. 1998)
under the assumption of ionisation equilibrium. This 
explicit finite differences code is based on a modified 
Leapfrog algorithm. Extensive tests and applications of 
the code can be found in the papers by
Schr\"oer et al. (1998),  Birk et al. (2001) and Birk
and Wiechen (2002).

Although the present study is motivated by the local 
information of the 
magnetised meteorite material in the solar system
we assume that our solar system is a standard case for 
normal stellar systems. Thus, in the following
we consider a simple protoplanetary accretion disk model 
consisting of 
electrons, protons, neutrals and charged dust grains 
using typical 
dynamical parameters for circumstellar disks around T 
Tauri stars
(Montmerle 1991; K\"onigl 1994).
In particular, we choose the following parameters:
$n_i = 1.01 n_d = 0.1 n_n=
10^{12}{\rm cm}^{-3}$, $z_d=1$, $m_d=100m_i$,
and for the temperatures $T_n=T_i=T_d=500$K. The ion-
neutral collision frequencies can be estimated as 
(e.g. Krall \& Trivelpiece 1986)
$\nu_{in} \approx \sigma(T_i) n_n \sqrt{k T_i/m_i}$
where $k$ is the Boltzmann constant and $\sigma$ denotes
the cross section. The latter depends on the ion temperature in a way
that results in a collision frequency that is independent on
the ion temperature (for more details on the collision theory see, e.g., 
Mitchner \& Kruger 1973).   
We use $\sigma=
5\cdot 10^{-15}{\rm cm}^2$ as constant approximation of sufficient
accuracy in our context. 
The ion-dust collision frequency is $\nu_{id}\approx 
4\sqrt{2\pi} n_d
z_d^2e^4 {\rm ln}(\lambda_D/a)/3 m_i^{1/2}(k T_i)^{3/2}$ 
(Benkadda et al. 1996)
where $\lambda_D=\sqrt{k T_d/4\pi n_dz_d^2e^2}$ 
denotes the dust Debye screening length. For the dust 
grain size
we choose $a=10^{-6}$cm. 
The actual choice of the grain size has no 
significant influence on the induction equation, 
since $a$ only enters in the logarithm of 
the formula for the ion-dust frequency and does not play 
an important role 
for the magnetic field generation due to the strong 
electron depletion ($z_d-n_i/n_d\approx 0$ in Eq.~10).
On the other hand, the momentum transfer between the dust 
and the neutral 
fluid components depends on the grain radius. However, 
the continuous
acceleration (accretion flow) guarantees a 
finite relative dust-neutral shear flow. Without this 
energy source
larger dust grains would result in faster saturation and 
weaker generated
magnetic fields.

Within a region of 10 AU
we consider an initially homogenous  protoplanetary
disk in which the dust grains as well as the ions are chosen 
to be in corotation within a region of about 1 AU
(Fig.1, upper plot). The corotation of the charged dust 
grains is caused by a stellar dipole field. The neutrals
do not couple directly to the magnetic field but via
momentum transfer with the charged particles.
Outside this corotation radius the dust is in Keplerian
motion (Li 1996). 
The neutral gas is assumed to be in Keplerian motion 
with the exception of a very inner region of 0.2 AU where 
the neutral gas 
is assumed to be in corotation to avoid singularities 
(Fig.1, lower plot). 
The typical velocities near the transition regions 
between corotation and 
Keplerian motion are about  $1\cdot 10^6$ cm s$^{-1}$ for 
the dust 
and $1.5\cdot 10^7$ cm s$^{-1}$ for the 
neutral gas, respectively.
\begin{figure}
\includegraphics[scale=0.6]{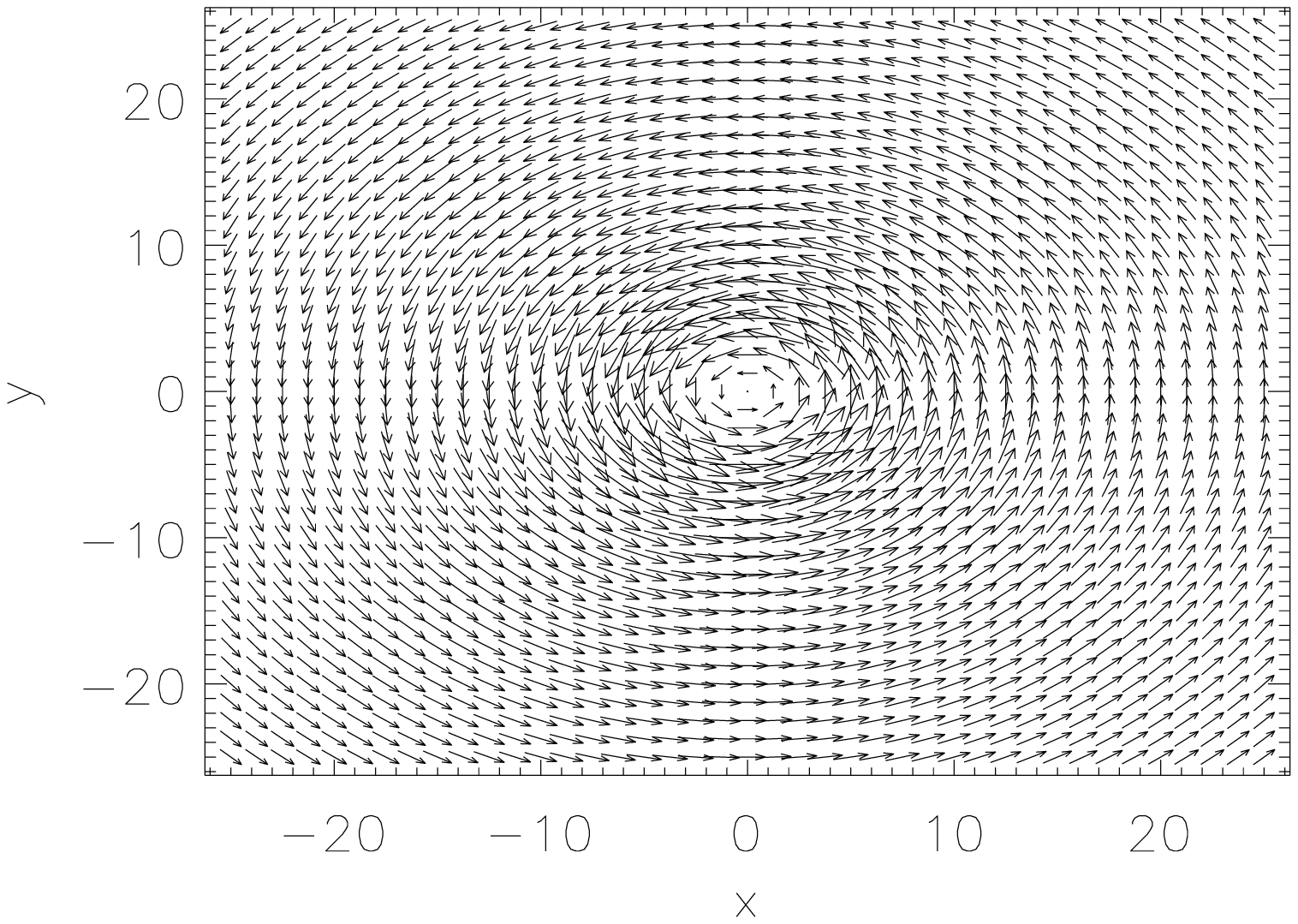}
\vskip-0.5truecm
\includegraphics[scale=0.6]{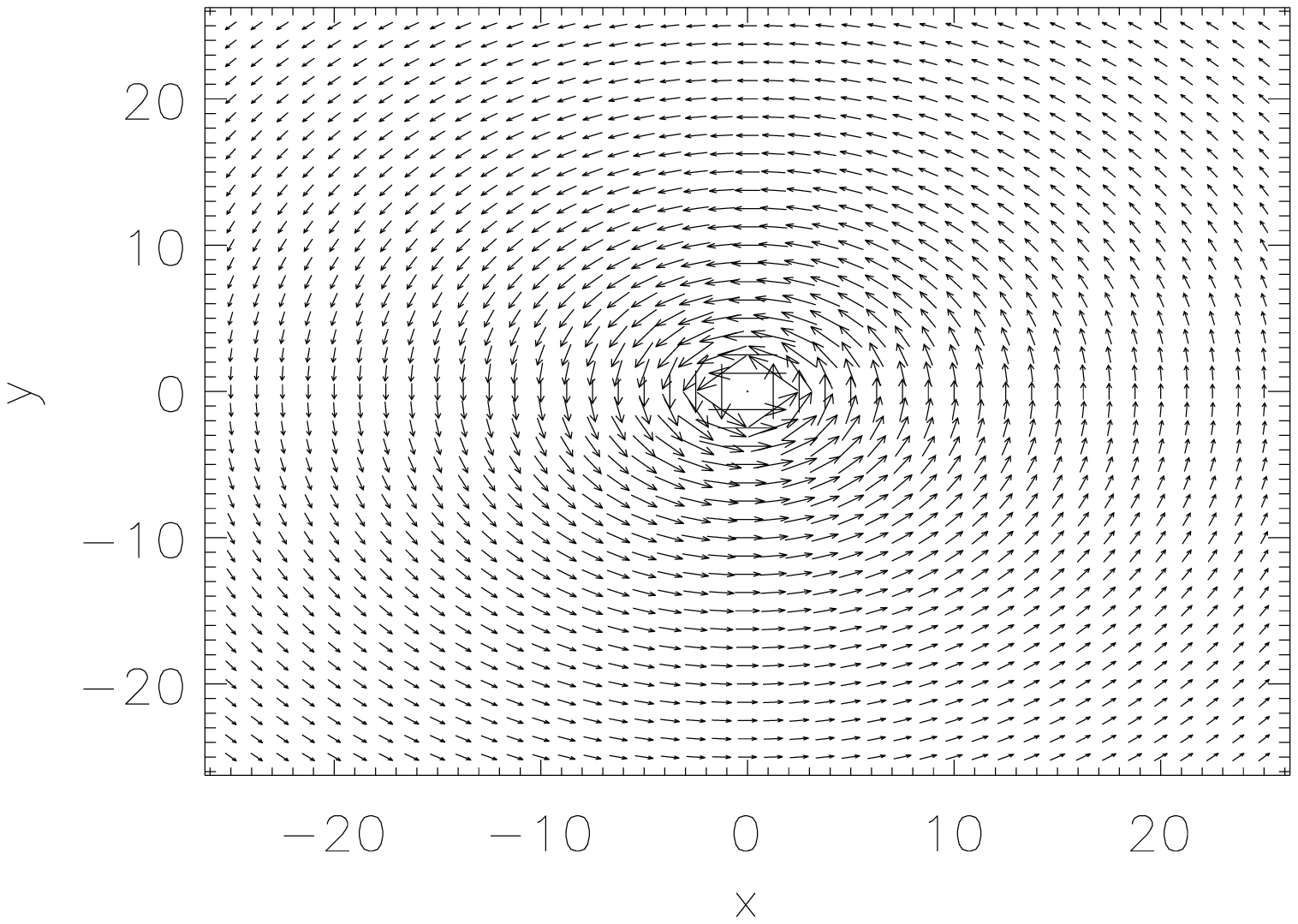}
\caption[]{The initial dust (upper plot) and neutral gas 
(lower plot)
velocity patterns. The velocity
field is initially invariant in the $z$-direction, i.e. 
perpendicular to
the disk. The spatial scales are given in units of 0.2 
AU.}
\label{fig1}
\end{figure}

The orders of magnitude of these velocities are in a 
rough agreement with a 
central mass object of about 2 solar masses, so that we 
include an external
gravitational force corresponding to such a central mass. 
The initial configuration is characterised by the 
immediate onset of idealised
accretion flow rather than to be in a stationary state. 
The gravitational field is the ultimate energy source for 
the generation of 
the magnetic field.
 
The simulation box is given by $x\in[-25, 25]$, 
$y\in[-25, 25]$ and $z\in[0, 10]$ in
units of 0.2 AU. 
The equidistant grid is resolved by $77\times 77
\times 31$ grid points. 
The boundary conditions are symmetric at all boundaries, 
i.e.
the values of the scalar quantities and vector components 
at the numerical 
boundaries correspond to the values at the respective 
last inner grid points.

As the upper plot of Fig.~2 indicates, 
a swirl-like dust flow in the central disk in the
$x$-$y$-plane has built up after $0.7$yr. 
The curl of the relative dust-neutral gas flow
(Fig.~2, lower plot) is sustained by the gravitational 
pull 
and results in an ongoing magnetisation of the disk.

\begin{figure}
\includegraphics[scale=0.6]{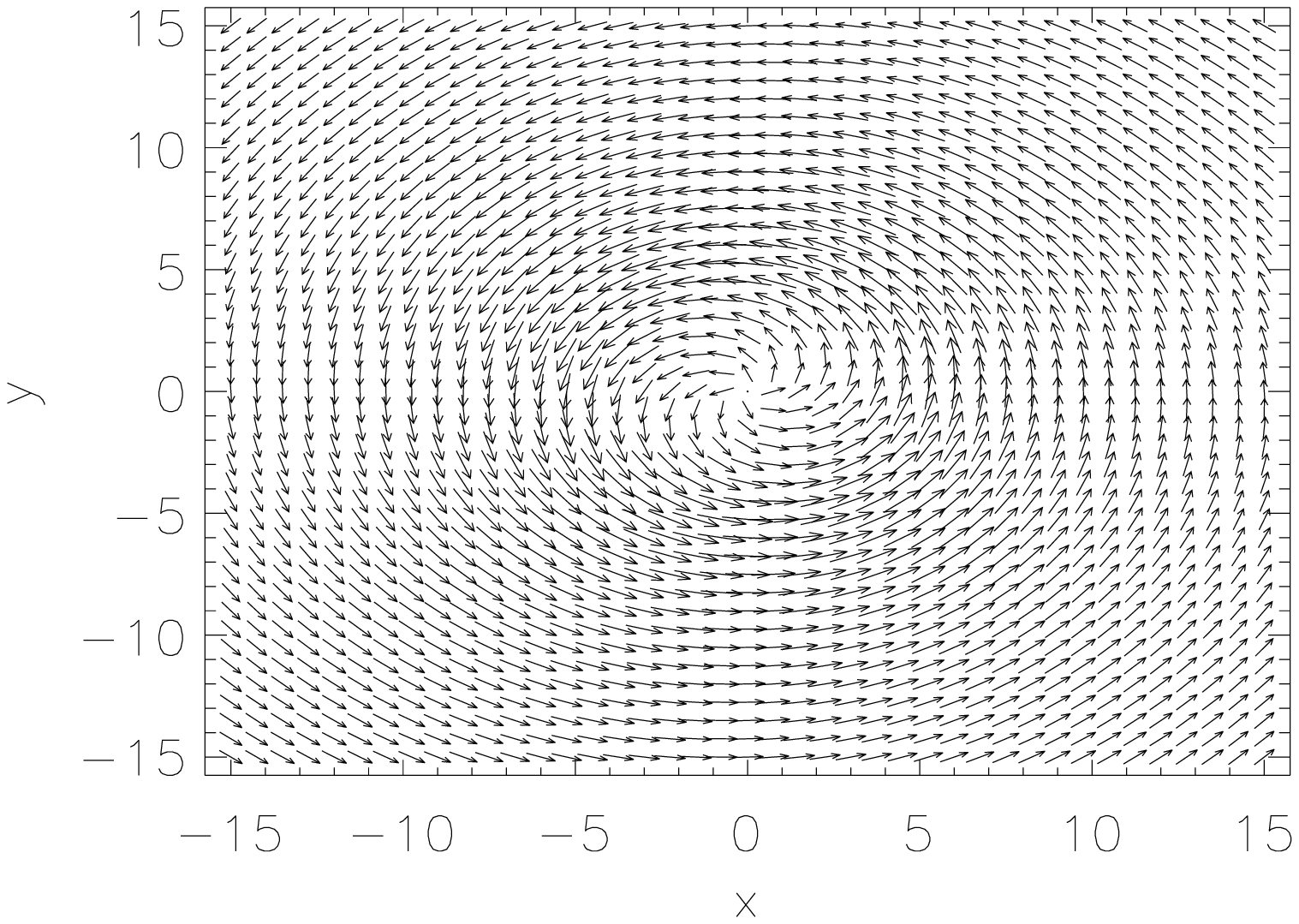}\end{figure}
\begin{figure}
\vskip-0.5truecm
\includegraphics[scale=0.6]{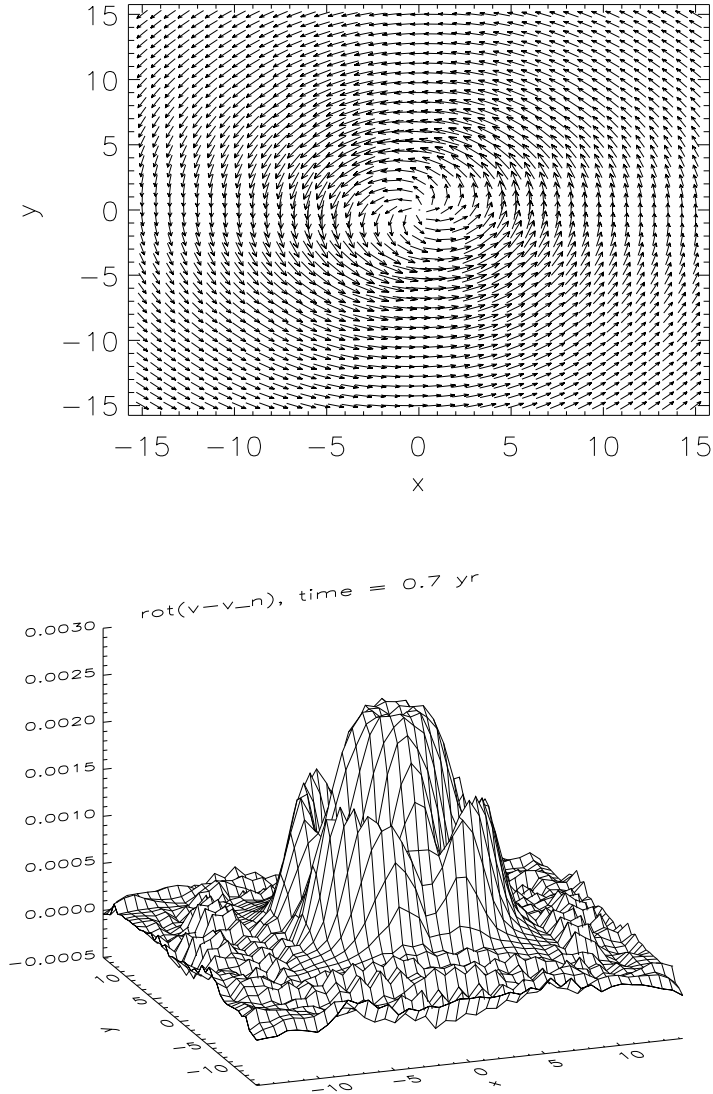}
\caption[]{The poloidal dust velocity patterns at the 
middle of the
disk (as far as the vertical extension goes) and the $z$-
component
of the curl of the 
relative dust-neutral gas flows after $t=0.7$yr. The 
amplitude is measured in
units of the initial amplitude of $v_{dx}$.
}\label{fig2}
\end{figure}

The magnetic field after $t=0.7$yr and $t=1.4$yr is shown 
in 
Figs.~3 and 4, respectively.
The strengths of the different field components are 
roughly comparable. 
After $t=0.7$yr, the amplitude of $B_z=0.8$G is 5 times 
the amplitudes 
of $B_x$ and $B_y$ and after $t=1.4$yr,
the factor is about 2 ($B_z=1.2$G). 
The field generation is caused not only by the poloidal
flows  that develop from the initial configuration but 
also by 
vertical relative shear flows arising in the dusty 
accretion disk.
\begin{figure}
\includegraphics[scale=0.6]{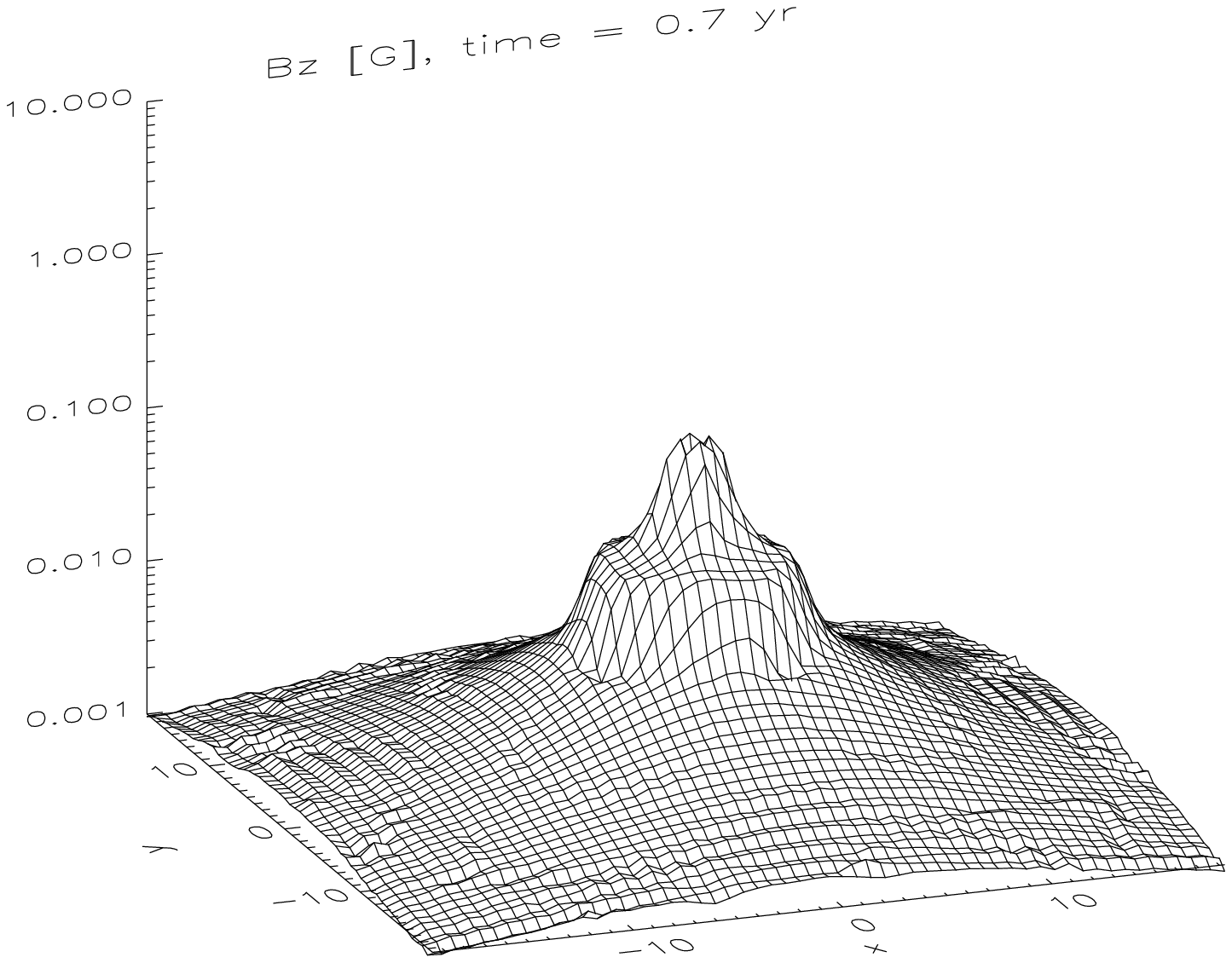}\end{figure}
\begin{figure}
\vskip-0.5truecm
\includegraphics[scale=0.6]{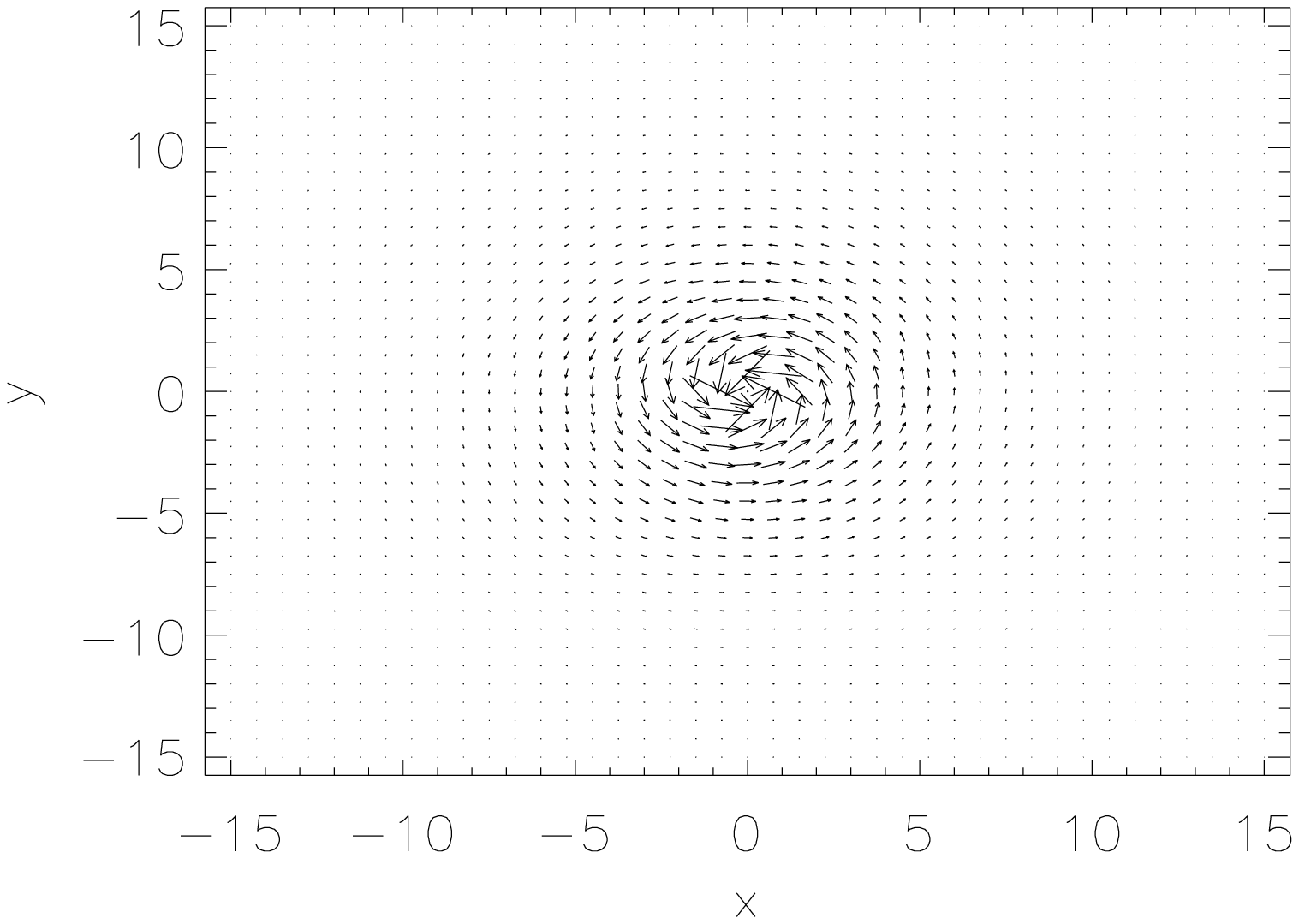}
\caption[]{The magnetic field after $t=0.7$yr.  
The upper plot shows the $z$-component
of the magnetic field and the lower one the poloidal 
magnetic field at $z=5$.}
\label{fig3}
\end{figure}
\begin{figure}
\includegraphics[scale=0.6]{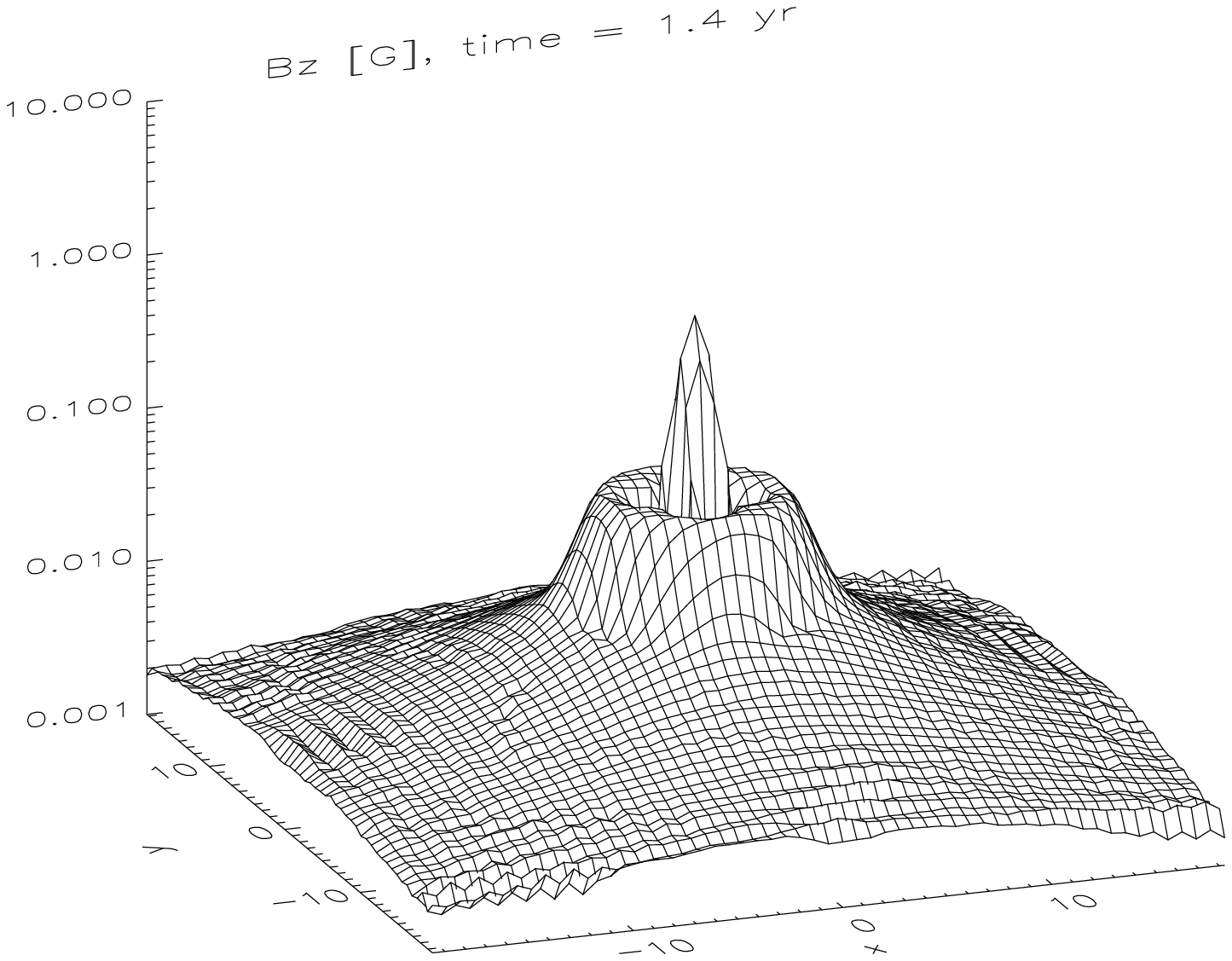}\end{figure}
\begin{figure}
\vskip-0.5truecm
\includegraphics[scale=0.6]{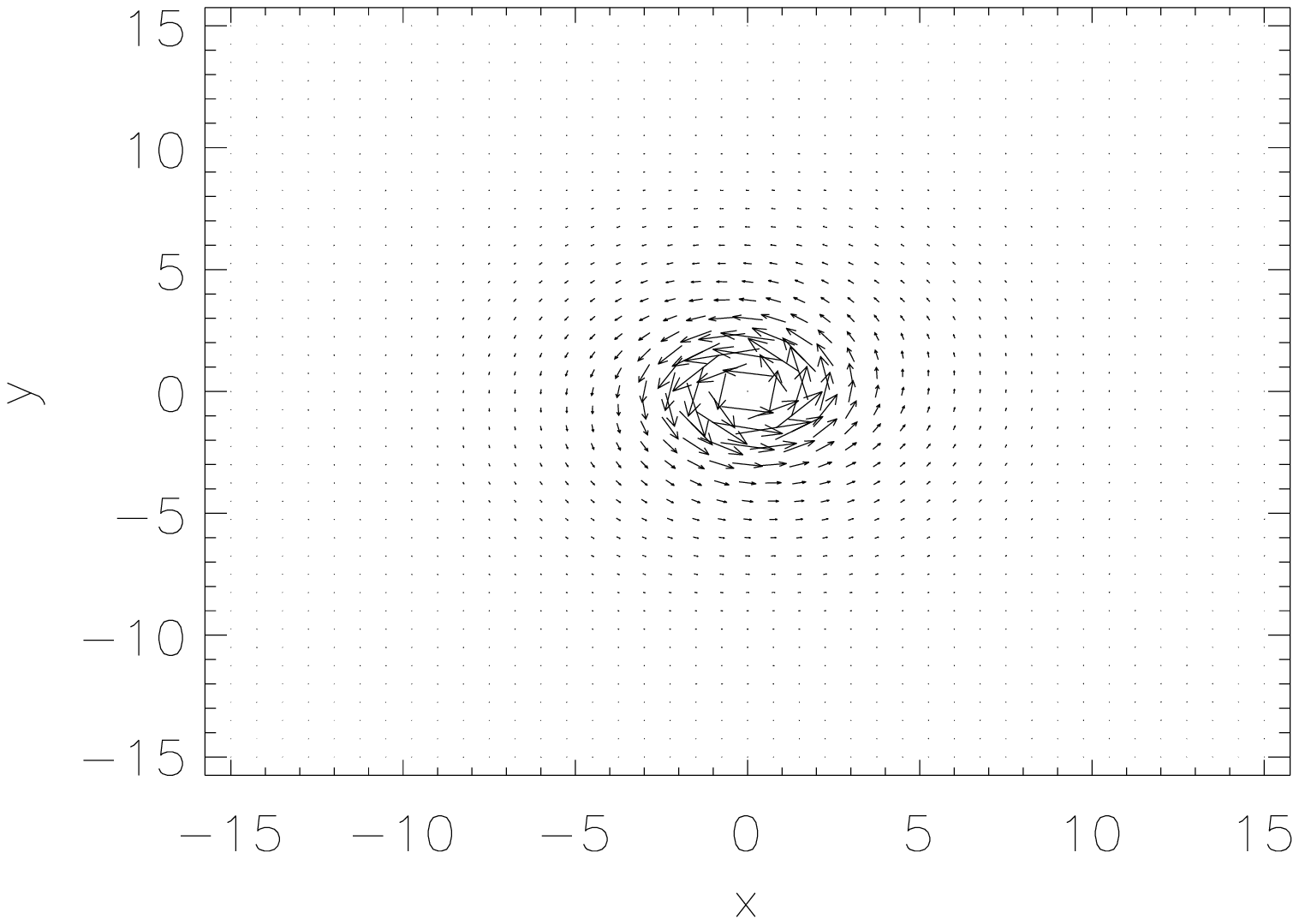}
\caption[]{The magnetic field after $t=1.4$yr. 
The upper plot shows the $z$-component
of the magnetic field and the lower one the poloidal 
magnetic field at $z=5$.}
\label{fig4}
\end{figure}

The magnetic energy density $W_B=B^2/8\pi$ at different 
heights 
of the disk are shown in Fig.~5. Obviously, the inner 
part of the disk  
is completely magnetised in all three dimensions.
\begin{figure}
\includegraphics[scale=0.6]{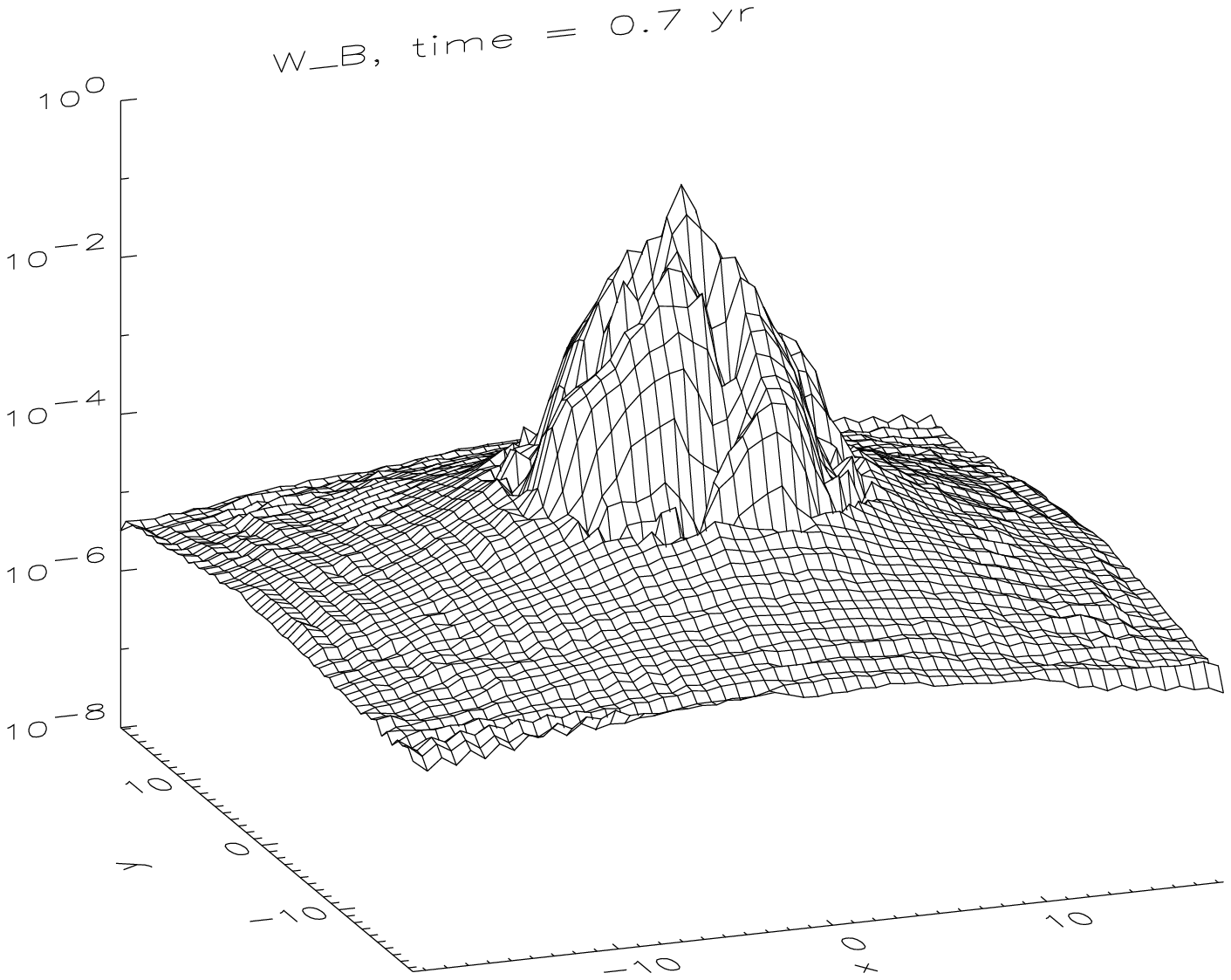}\end{figure}
\begin{figure}
\vskip-0.5truecm
\includegraphics[scale=0.6]{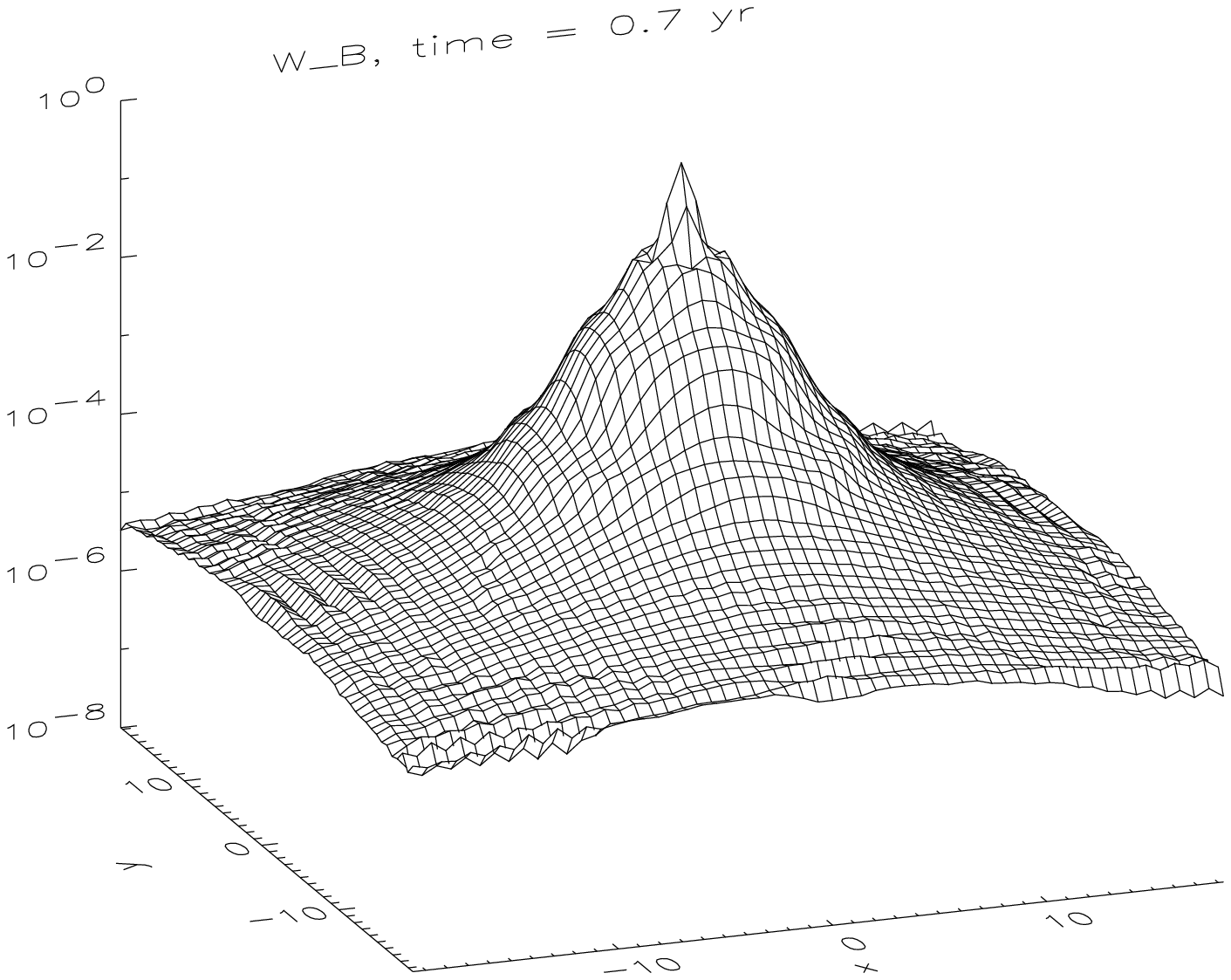}
\vskip-0.5truecm
\includegraphics[scale=0.6]{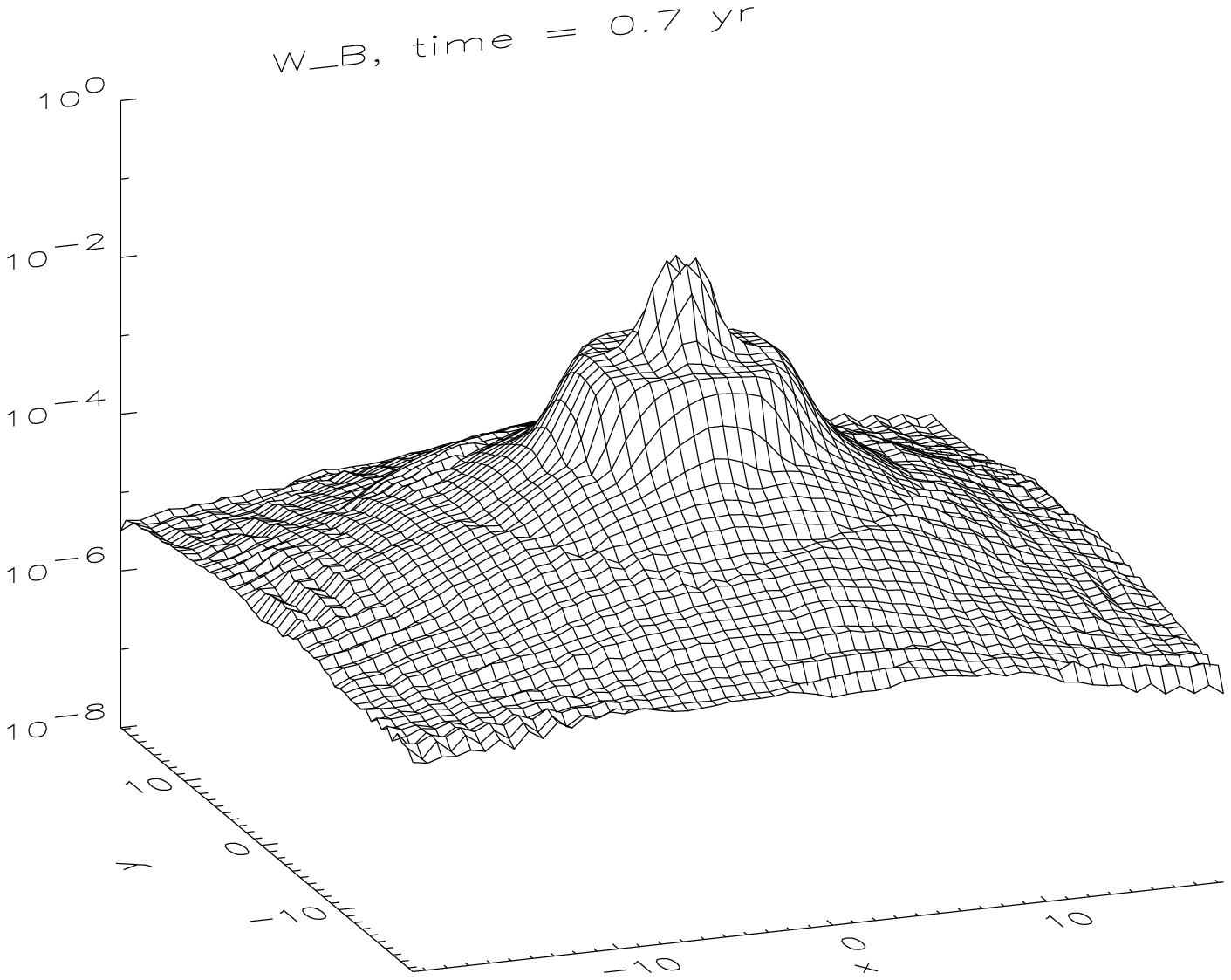}
\caption[]{The magnetic field energy density ($W_B$) 
 at different heights
($z=1.4$, upper plot; $z=5$, middle plot; $z=8.6$ lower 
plot)
of the protoplanetary disk.} 
\label{fig5}
\end{figure}

Since the ultimate source of the self-magnetisation is the
stellar gravitation the process lasts as long as the accretion flow
in the circumstellar disk 
is present. The numerical simulation, however, fails when the central
density gradient becomes too steep (after about 200 dynamical times). 
This is a consequence of our simulation
approach that does not properly deal with the accretion of matter 
(loss of plasma to the stellar object) and the formation of bipolar jets. 
A global multi-fluid model including all relevant physics
is clearly beyond the scope of the present state of the art.
 
\section{Discussion}

We studied the magnetisation of protoplanetary disks that consist
of dusty plasmas.
Different from previous studies on the magnetisation of the environment
of collapsing molecular cloud cores \citep{cio98, des01}
we dwell on the magnetic field generation in fully developed
accretion disks.
The process is driven by accretion, i.e. the ultimate source of energy for the 
magnetic field is the stellar gravitational field, it results in 
an ordered poloidal magnetic field of 
some Gauss. It is a very fast and efficient mechanism
of field generation. Also it is a process of permanent magnetisation as 
long as some accretion flow is present.

The numerical simulations show that magnetic fields are 
generated in circumstellar dusty
disks by macroscopic relative shear motions between the 
different species.
This self-magnetisation mechanism 
that works in addition to the 
formation of turbulent magnetic fields due to the 
interaction of the stellar
field with a turbulent accretion disk (Bardou \& 
Heyvaerts 1996). 

We note that in our studies we did not include a 
stellar magnetic field
explicitly. Such a field should give rise to a (sheared) 
wind component
close to the star and thus, possibly to the further  
magnetisation of a significant fraction of
the magnetosphere. In this contribution we concentrated 
on the generation of 
magnetic fields in the accretion disks. Further studies 
that include stellar
fields are a promising task for the future.

In our simulations we did not include a gap between the 
central stellar 
object and the disk for technical reasons.  
However, since we were interested in the in 
situ magnetisation of the disk material this idealisation 
does not influence 
our findings.

A further idealisation of our study is the initial 
density
homogeneity of the disk. However, in the  course of the 
dynamical evolution
radial profiles form in all fluid components due to the 
external gravitation.
More realistic density profiles perpendicular to the disk 
would demand for 
the inclusion of self-gravitation. The inclusion of this 
further 
energy source that also may contribute to the self-
magnetisation of the disk
is beyond the scope of the present investigations but 
could be of 
interest for future numerical modelling. Parameter 
studies, however,  
have shown that the efficiency of the process of
magnetisation is not significantly influenced by the 
variation of
the initial number densities over a wide parameter range.  

Given the uncertainties of the actual plasma parameters 
the high efficiency
of the discussed magnetisation guarantees 
that protoplanetary disks are expected
to generate their own magnetic fields of the order of 0.1 
-- 1 Gauss.
In the presence of a small radial magnetic field 
component the 
Balbus-Hawley shear instability (Balbus \& Hawley 1991)
can, in principle, result in a faster than linear growth
of the poloidal magnetic field component
provided the appropriate onset criterion is fulfilled.
This criterion, however, is not known for multifluid 
systems.
On the other hand, localized reconnection
may result in a redistribution of magnetic flux.
Investigations on this process are beyond the scope of 
the present 
contribution and will be considered in further more 
detailed work.

If circumstellar disks carry their own significant 
magnetic fields, dynamic phenomena in the coupled star-disk system are 
rather complex.
The disk field should influence characteristic phenomena 
as, e.g., accretion, 
winds and jets and magnetic activity, significantly.

\bsp

\label{lastpage}

\end{document}